\documentclass[journal]{IEEEtran}
\pdfoutput=1
\usepackage[noadjust]{cite}
\usepackage{amsmath}
\usepackage{array}
\usepackage{mdwmath}
\usepackage{amssymb}
\usepackage{algorithm}
\usepackage[noend]{algpseudocode}
\usepackage{eqparbox}
\usepackage{stfloats}
\usepackage{url}
\usepackage[pdftex]{graphicx}
\graphicspath{{../pdf/}{../jpeg/}}
\DeclareGraphicsExtensions{.pdf,.jpeg,.png}
\usepackage{epstopdf}
\usepackage{amsfonts}

\usepackage{color}

\usepackage{tikz}
\usepackage{pgfplots}
\usepackage{subcaption}

\newtheorem{theorem}{Theorem}
\newtheorem{remark}{Remark}
\newtheorem{corollary}{Corollary}

\begin{document}

\title{Opportunistic Beamforming using an Intelligent Reflecting Surface Without Instantaneous CSI}
\author{\hspace{-.02in}Qurrat-Ul-Ain\hspace{-.02in}~Nadeem,\hspace{-.02in}~\IEEEmembership{Member,~IEEE,} Anas\hspace{-.02in}~Chaaban,\hspace{-.02in}~\IEEEmembership{Senior \hspace{-.02in} Member,\hspace{-.02in}~IEEE,} \hspace{-.02in} and \hspace{-.02in}
        M{\'e}rouane~\hspace{-.02in}Debbah,\hspace{-.02in}~\IEEEmembership{Fellow,\hspace{-.01in}~IEEE} 
\thanks{Q.-U.-A. Nadeem and A. Chaaban are with School of Engineering, The University of British Columbia, Kelowna, Canada.  (e-mail:  \{qurrat.nadeem, anas.chaaban\}@ubc.ca)}
\thanks{M. Debbah is  with  Universit{\'e} Paris-Saclay, CNRS, CentraleSup{\'e}lec, 91190, Gif-sur-Yvette, France (e-mail: merouane.debbah@centralesupelec.fr).}
}

\markboth{}%
{Shell \MakeLowercase{\textit{et al.}}: Bare Demo of IEEEtran.cls for Journals}
\maketitle

\begin{abstract}

While intelligent reflecting surface (IRS) assisted wireless communication has emerged as an important research paradigm, channel state information (CSI) acquisition remains a critical challenge to design the IRS phase-shifts and yield the promised coherent beamforming gains. In this paper, we propose an IRS-assisted opportunistic beamforming (OBF) scheme under proportional fair scheduling, which does not require instantaneous CSI to design the IRS parameters. In a slow-fading environment, we show that with only random rotations at the IRS, the proposed scheme can capitalize on the multi-user (MU)-diversity effect to approach the performance of coherent beamforming as the number of users grows large. Next we study the sum-rate scaling of IRS-assisted OBF in the correlated Rayleigh fast fading environment under  a deterministic  beamforming scheme that results in a considerable sum-rate improvement.

\end{abstract}
\vspace{-.05in}

\vspace{-.15in}
\section{Introduction}

Intelligent reflecting surface (IRS) can help realize reconfigurable propagation channels between the base station (BS) and the users. An IRS is abstracted as an array of passive reflecting elements, each of which can independently introduce a phase-shift onto the impinging electromagnetic waves to achieve different communication goals, for example: maximize the system's energy efficiency \cite{huang} or minimize the transmit power \cite{LIS, robust} subject to quality of service constraints, and maximize the minimum rate \cite{annie} subject to transmit power constraints.

Existing works yield coherent beamforming gains by optimizing the IRS phase-shifts under the assumption of perfect channel state information (CSI), which is highly impractical given the radio limitations of the passive IRS. In fact, the recently developed channel estimation protocols require the training time to grow proportionally with the number of IRS elements, thus hampering most of the expected reflect beamforming gains \cite{annie_OJ, broadband}. Moreover, optimizing IRS at the coherence time-scale level increases the system complexity.

Motivated by these challenges, we study an IRS-assisted single-input single-output (SISO) broadcast channel (BC), in which the IRS elements introduce random or deterministic phase rotations without requiring instantaneous CSI.  The average sum-rate capacity of the SISO BC is achieved by opportunistic scheduling (OS), which schedules at one time the user with the largest signal-to-noise ratio (SNR) \cite{TDMA1,dumb}.  MU-diversity gains then arise because in a system with many users whose channels fade independently, there is likely to be a user at each time whose SNR is near its peak. However, these gains are severely limited when channels fade slowly \cite{dumb, slowfad}. For such scenarios, Viswanath \textit{et al.} \cite{dumb} proposed an opportunistic beamforming (OBF) scheme, where multiple BS antennas transmit weighted replicas of the same signal to induce temporal channel variations and  improve the sum-rate. 

Instead of using multiple active  BS antennas for OBF, we propose to utilize a passive IRS in the SISO BC, where the IRS  elements induce time-varying random phase rotations. Each user feeds back its downlink SNR and the BS employs proportional fair (PF) scheduling, which captures most of the MU-diversity gain promised by OS while maintaining user fairness \cite{dumb}. We present an asymptotic analysis of the sum-rate under slow fading, which reveals that the IRS-assisted OBF scheme can capitalize on the artificially induced MU-diversity effect to approach the coherent beamforming performance as the number of users increases.  We also study  the sum-rate scaling for the correlated Rayleigh fast fading scenario under  a deterministic design for the IRS phase-shifts and show significant sum-rate gains without requiring instantaneous CSI. 

To this end, we point that the results in this work can be extended in the future to the multi-antenna BC under random beamforming \cite{TDMA1,tareq}, where multiple orthonormal beams are transmitted from the BS and on each beam the strongest user is served, while the IRS elements induce random phase rotations. We also remark that the only other works that study the random rotations-based IRS scheme are \cite{rotate1} and \cite{dumb_me}, where the former studies its impact on the outage probability and energy efficiency of  a  point-to-point SISO system, while the latter studies its effect on the sum-rate scaling of  SISO Rayleigh and Rician fading BCs.  Random rotations-based IRS scheme is also inspired from the rotate-and-forward protocol in   \cite{rotate} that converts a slow-fading relay channel into a time-varying channel using time-varying random rotations.

The rest of the paper is organized as follows. Sec. \ref{Sec:Model_and_OBF} introduces IRS-assisted OBF, Sec. \ref{Sec:Asym_Analysis_RS_OBF} presents the asymptotic sum-rate analysis in slow and fast fading channels, Sec. \ref{Sec:Simulations} provides simulation results and Sec. V concludes the paper.
\vspace{-.12in}

\section{System Model}\label{Sec:Model_and_OBF}

\subsection{Transmission Model}

We consider the downlink communication between a single-antenna BS and $K$ single-antenna users over block-fading channels, $h_k(t)\in\mathbb{C}$, which remain constant during a frame $t$ of length $T$ symbols corresponding to the coherence interval. The received signal $\mathbf{y}_{k}(t) \in \mathbb{C}^{T\times 1}$ at user $k$ in frame $t$ is 
\begin{align}
\label{t_model}
&\mathbf{y}_{k}(t)=h_{k}(t)\mathbf{s}(t)+\mathbf{n}_{k}(t),
\end{align}
where $\mathbf{s}(t)\in \mathbb{C}^{T\times 1}$ is the vector of $T$ transmitted  symbols from the BS in frame $t$ and $\mathbf{n}_{k}(t)\in \mathbb{C}^{T\times 1}$ is the noise vector at user $k$ in frame $t$ distributed as $\mathcal{CN}(\mathbf{0}, \sigma^2 \mathbf{I}_{T})$, where $\sigma^2$ is the noise variance.  We assume  that the Tx power level, denoted as $P$, is fixed at at all times and therefore the Tx signal vector $\mathbf{s}(t)$ must satisfy the power constraint $\mathbb{E}[||\mathbf{s}(t)||^2]=PT$.

In this SISO BC with only SNR CSI, the sum-capacity is achieved by OS, wherein the BS transmits to the user with the highest SNR  \cite{dumb,TDMA1}. The scheduled user in frame $t$ is $\hat{k}(t)= \underset{{k\in\{1,\ldots, K\}}}{\arg\max}\gamma_{k}(t)$, where $\gamma_{k}(t) =\frac{P|h_{k}(t)|^2}{\sigma^2}$ and $R_k(t)=\log_2(1+\gamma_{k}(t))$ are the SNR and requested rate of user $k$. The maximum SNR is $\gamma_{\hat{k}(t)}(t)$ and average sum-rate capacity is
\begin{align}
\label{AR}
R^{(K)}= \mathbb{E}[\log_2(1+\gamma_{\hat{k}(t)}(t))],
\end{align}
where the expectation is over $(h_{1}(t), \ldots, h_{K}(t))$.

When the users’ fading statistics are the same, OS maximizes not only the sum-capacity but also the fairness among users. In reality, the users will have different path losses resulting in weak users to almost never be scheduled. To address this issue while exploiting the MU-diversity gains promised by OS, PF scheduling was proposed to keep track of the average throughput of each user $T_{k}(t)$ in a past window of length $t_c$ frames and  schedule the user in frame $t$ as \vspace{-.04in}
\begin{align}
\label{PF}
\hat{k}(t)= \underset{{k\in\{1,\ldots, K\}}}{\arg\max}R_{k}(t)/T_k(t).
\end{align} 
The average sum-rate is still expressed as \eqref{AR} with $\hat{k}(t)$ given in \eqref{PF}. When the channels undergo  fast fading, $R^{(K)}$ increases with $K$ due to the MU-diversity effect while it stays constant under slow-fading \cite{dumb}.  Next we propose IRS-assisted OBF that yields MU-diversity gains under both slow and fast fading.

\vspace{-.1in}
\subsection{IRS-Assisted OBF}

\begin{figure}
\centering
\includegraphics[scale=.21]{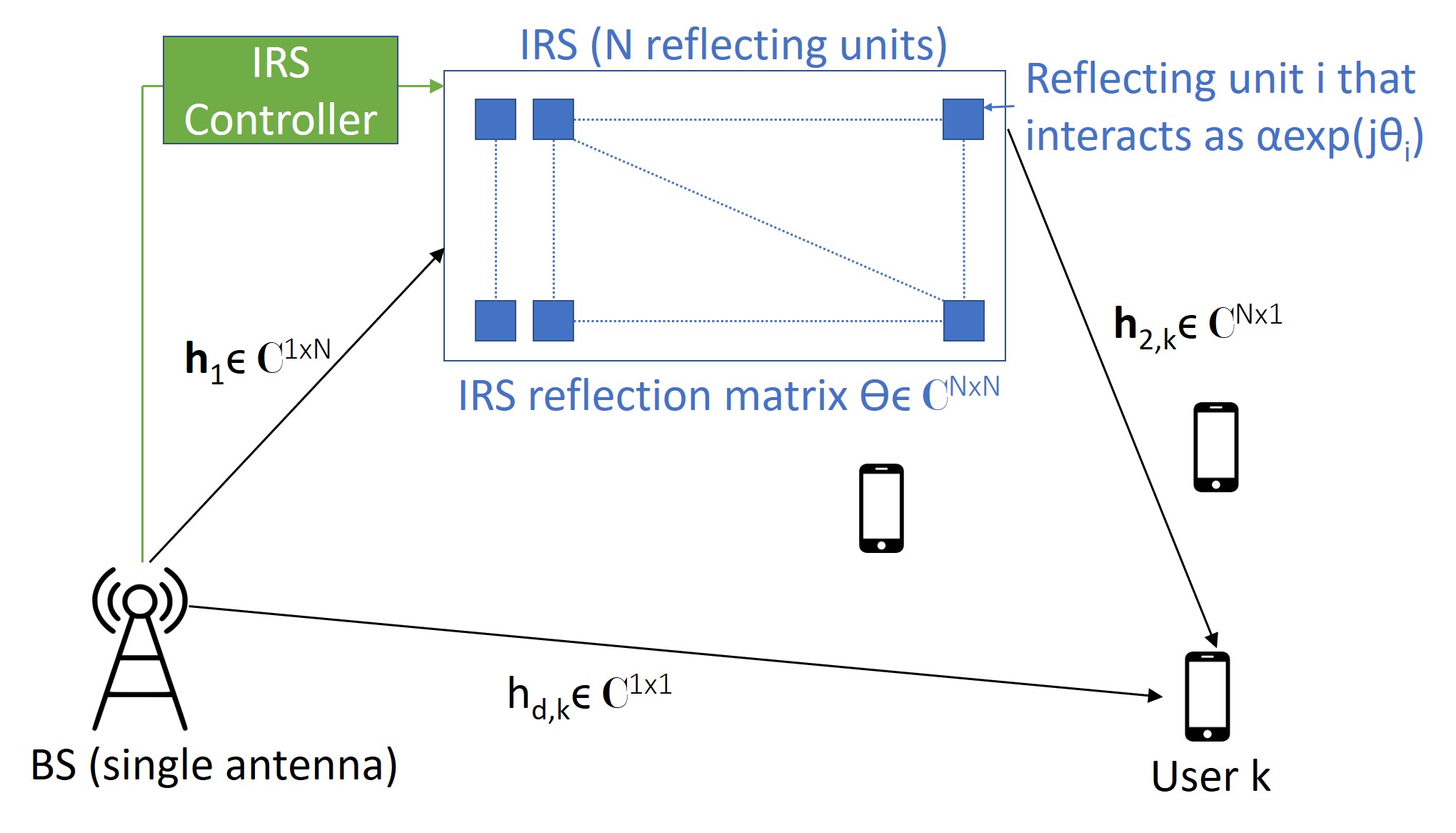}
\caption{IRS-assisted OBF system model.}
\label{LIS_model}
\end{figure}

An IRS composed of $N$ passive reflecting elements is installed in the SISO BC to assist the BS in communicating with the users as shown in Fig. \ref{LIS_model}. The IRS elements introduce random phase shifts onto the incoming waves in each coherence interval. The channel $h_{k}(t)$ in frame $t$ is given as 
\begin{align}
\label{ch_IRS}
h_{k}(t)&=\sqrt{\beta_{r,k}}\mathbf{h}_1 \boldsymbol{\Theta}(t) \mathbf{h}_{2,k}(t)+\sqrt{\beta_{d,k}}h_{d,k}(t),\\
\label{ch_IRS2}
&=\sqrt{\beta_{r,k}}\mathbf{v}(t)^T \text{diag}(\mathbf{h}_{1}) \mathbf{h}_{2,k}(t)+\sqrt{\beta_{d,k}}h_{d,k}(t),
\end{align}
where $\beta_{r,k}$ and $\beta_{d,k}$ are the signal attenuation factors for the IRS-assisted and direct links respectively, $\mathbf{h}_{1}\in\mathbb{C}^{1\times N}$ is the BS-IRS channel vector, $\mathbf{h}_{2,k}(t) \in \mathbb{C}^{N\times 1}$ is the IRS-user $k$ channel vector and $h_{d,k}(t) \in \mathbb{C}$ is the direct BS-user $k$ channel. Moreover $\boldsymbol{\Theta}(t)=\alpha\text{diag}\{e^{j\theta_{1}(t)}, \ldots,e^{j\theta_{N}(t)}\} \in \mathbb{C}^{N\times N}$ is a diagonal matrix representing the response of the IRS, where $\alpha\in [0,1]$ is the fixed amplitude reflection coefficient and $\theta_n(t)\in [0,2\pi]$ is the phase shift applied by $n$-th element. The specific way the $\theta_n$'s are generated is studied in the next section and will not require instantaneous CSI. The second reformulation in \eqref{ch_IRS2} has $\mathbf{v}(t)=\alpha [e^{j\theta_{1}(t)}, \ldots, e^{j\theta_{N}(t)}]^T$.  

We assume the BS-IRS channel to be LoS similar to many other works on this subject \cite{annie, annie_OJ, broadband}. The assumption is practical because both the BS and IRS are generally elevated high and will therefore have a very few ground structures around to block or reflect the electromagnetic waves. Moreover, any NLoS paths in the BS-IRS channel are expected to experience a much higher path loss in next generation communication systems and can therefore be neglected as compared to the LoS path \cite{broadband}. The $n$-th component of $\mathbf{h}_1$ is given as $h_{1,n}=e^{j \vartheta_{h_1,n}}$, where $\vartheta_{h_1,n}=2\pi (n-1) d\sin {\vartheta}_n$ \cite{annie}, ${\vartheta}_n$ is the LoS angle to IRS element $n$, and $d$ is the inter-element separation. 

As outlined in Sec. \ref{Sec:Model_and_OBF}-A, the users feedback their SNRs $\gamma_{k}(t)$, and the BS schedules the user with the largest $\frac{R_k(t)}{T_{k}(t)}$ in frame $t$. The only  feedback required from each user under the IRS-assisted OBF scheme is therefore just its SNR value, while no instantaneous CSI is needed to tune the IRS phase shifts as we will show in the next section. Next we will study the average sum-rate in \eqref{AR} under this scheme for both slow-fading and correlated Rayleigh fast fading channels. 

\vspace{-.1in}
\section{Asymptotic Analysis of the Sum-Rate}\label{Sec:Asym_Analysis_RS_OBF}

We consider the large $K$ regime in our analysis, which is very relevant given the massive connectivity promised by 5G. 

\vspace{-.1in}
\subsection{Slow Fading}\label{Sec:Asym_Analysis_RS_OBF_Ray}

We first consider the case of slow fading where the channel gains of each user remain constant, i.e. $\mathbf{h}_{2,k}(t)=\mathbf{h}_{2,k}$, ${h}_{d,k}(t)=h_{d,k}$, $\forall t$ (practically this means for all $t$ over the latency time scale of interest). The received SNR at each user will remain constant if no IRS is used and no MU-diversity gain will be exploited. Under the proposed scheme, however, the overall channel $h_{k}(t)$ in \eqref{ch_IRS} still varies over time due to $\boldsymbol{\Theta}(t)$ and the sum-rate can be improved through MU-diversity.

First, we present the maximum achievable rate for each user under coherent beamforming at the IRS with full perfect CSI, which will serve as a benchmark for comparison.
\begin{theorem}\label{Thm:RS_Coh}
The maximum rate achieved by user $k$ under coherent beamforming at the IRS is 
\begin{align}
\label{RBF}
&R_k^{BF}=\log_2\Big(1+\frac{P}{\sigma^2} \|\alpha \sqrt{\beta_{r,k}} \sum_{n=1}^N |{h}_{1,n}| |{h_{2,k,n}}|\nonumber \\
&\times \exp(j\angle h_{d,k})+\sqrt{\beta_{d,k}} h_{d,k}\|^2\Big),
\end{align}
where ${h}_{1,n}$ and ${h}_{2,k,n}$ are the $n^{th}$ elements of $\mathbf{h}_{1}$ and $\mathbf{h}_{2,k}$ respectively. This is achieved when $\theta_n(t)$ is set as
\begin{align}
\label{conf}
&\theta^{BF}_{n,k}=\angle h_{d,k}- \angle({h}_{1,n}+{h}_{2,k,n}), \hspace{.05in} n=1,\dots, N.
\end{align}
\end{theorem}
\begin{IEEEproof}
The proof follows from expressing $h_k(t)$ in \eqref{ch_IRS} as $h_k(t)= \alpha \sqrt{\beta_{r,k}}\sum_{n=1}^N \exp(j\theta_n(t)) |h_{1,n}||h_{2,k,n}|\exp(j\angle(h_{1,n}+h_{2,k,n}))+\sqrt{\beta_{d,k}} |h_{d,k}|\exp(j\angle h_{d,k})$ and noting that $|h_k(t)|^2$ is maximized using the beamforming configuration in \eqref{conf}. 
\end{IEEEproof}

To achieve coherent beamforming to a user, the BS will require full CSI of $h_{d,k}$ and $\mathbf{h}_{2,k}$, the latter being extremely difficult to obtain given the IRS is passive \cite{annie_OJ}. In the following, we show that as the number of users increases, the average rate of each user where the users just feedback the overall SNR and the IRS phases are randomly varied from a specified distribution, approaches that under coherent beamforming. 
 
Denote by $T_k^{(K)}$ the long-term average rate of user $k$ in a system with $K$ users when PF scheduling with infinite window ($t_c=\infty$) is used. We obtain the following results for $T_k^{(K)}$ and $R^{(K)}$ under IRS-assisted OBF as $K$ grows large.

\begin{theorem}\label{Thm:slow} 
Suppose the slow fading states of users are i.i.d. and discrete, and the joint stationary ergodic distribution of $(\theta_1(t), \dots, \theta_N(t))$ is the same as that of \vspace{-.04in}
\begin{align}
\label{eq_imp}
&(\theta_{1,k}^{BF}, \dots, \theta_{N,k}^{BF}),
\end{align}
for the slow fading state of any individual user $k$, where $\theta_{n,k}^{BF}$ is defined in \eqref{conf}. Then, almost surely, we have \vspace{-.05in}
\begin{align}
\label{Theorem1}
&\underset{K\to \infty}{\lim} KT_k^{(K)}=R_k^{BF},  \hspace{.02in} \forall k,
\end{align} 
\end{theorem}
where $R_k^{BF}$ is the rate under coherent beamforming in \eqref{RBF}.
\begin{IEEEproof}
See Appendix \ref{Proof_Thm_slow}. 
\end{IEEEproof}

\begin{corollary} 
An asymptotic approximation of the average sum-rate in \eqref{AR} under the setting of Theorem \ref{Thm:slow} is obtained as \vspace{-.09in}
\begin{align}
\label{Thm1_1}
\underset{K\to \infty}{\lim} R^{(K)}=\frac{1}{K}\sum_{k=1}^K R_k^{BF}.
\end{align} 
\end{corollary}
\begin{IEEEproof}
The average sum-rate $R^{(K)}=\sum_{k=1}^K T_k^{(K)}$ for PF scheduling with infinite window and use \eqref{Theorem1}. 
\end{IEEEproof}

This result implies that for large $K$, with high probability the PF algorithm always schedules the users when they are in their beamforming configurations. Moreover, it allocates equal amount of time to each user as signified by the $\frac{1}{K}$ factor in \eqref{Thm1_1}. The sum-rate performance increases with $K$ despite the slow-fading nature of channels due to the artificially introduced channel fluctuations through the IRS and approaches the coherent beamforming performance. This is done using only random IRS phase shifts drawn from the  stationary distribution specified by  \eqref{eq_imp}, without requiring instantaneous CSI. 

\vspace{-.1in}
\subsection{Correlated Rayleigh  Fading}
\vspace{-.02in}

While IRS-assisted OBF increases the rate of channel variation in slow fading channels, it will not have the same effect under fast fading. However, better MU-diversity gains can still be exploited if IRS increases the dynamic range of distribution of $h_k(t)$ in \eqref{ch_IRS}. To study this, we analyze  the asymptotic behaviour of \eqref{AR} for the scenario where $h_{d,k}\sim \mathcal{CN}(0,1)$ and \vspace{-.07in}
\begin{align}
\label{ch_corr}
&\mathbf{h}_{2,k}(t)=\mathbf{R}_k^{\frac{1}{2}} \mathbf{b}_{k}(t),
\end{align}
where $\mathbf{b}_{k}(t)\sim \mathcal{CN}(\mathbf{0},\mathbf{I}_N)$ and $\mathbf{R}_k$ is the $N\times N$ correlation matrix at the IRS with $\text{trace}(\mathbf{R}_k) = N$. For this fading model, we rely on tools from extreme value theory to study the asymptotic scaling of $\underset{k}{\text{max }}{\gamma_k}$ in \eqref{AR} in the limit of a large number of users, under the assumption that the fading statistics of all users are identical, i.e. $\mathbf{R}_k=\mathbf{R}$, $\beta_{r,k}=\beta_r$, $\beta_{d,k}=\beta_{d}$ $\forall k$\footnote{This assumption is for analytical tractability since it will make the users' channels (and SNRs) i.i.d., enabling the application of results from extreme value theory. However, one could argue as \cite{dumb, tareq}  that similar multi-user diversity gains will be observed when the fading statistics are non-identical.}\cite{dumb, TDMA1,tareq}. Under this setting, PF scheduling reduces to OS, i.e. transmit to user with highest $|h_k(t)|^2$ during frame $t$. 

The overall channel $h_{k}(t)$ in \eqref{ch_IRS} is given as $h_{k}(t)=\sqrt{\beta_r}\mathbf{v}(t)^T \bar{\mathbf{h}}_{k}(t)+\sqrt{\beta_{d}} h_{d,k}(t)$, where $\bar{\mathbf{h}}_{k}(t)= \text{diag}(\mathbf{h}_{1}) \mathbf{h}_{2,k}(t)$ and is distributed as $\mathcal{CN}(\mathbf{0},\bar{\mathbf{R}})$ with $\bar{\mathbf{R}}=\text{diag}(\mathbf{h}_{1})\mathbf{R}\text{diag}(\mathbf{h}_{1}^H)$ and $\text{trace}(\bar{\mathbf{R}})=N$. Dropping the time-index for simplicity,  the sum-capacity in \eqref{AR} is given as
\begin{align}
\label{AR2}
R^{(K)}=\underset{f(\mathbf{v})}{\text{max } }\mathbb{E}_{\mathbf{v}}[\mathbb{E}_{h_{1},\dots, h_K|\mathbf{v}}[ \log_2(1+\underset{k}{\max} \gamma_{k})]],
\end{align}
where $\mathbb{E}_{x|y}$ is the conditional expectation of $x$ given $y$. Note that the channel ${h}_{k}$ for given $\mathbf{v}$ is distributed as 
\begin{align}
\label{hh}
&h_{k}|\mathbf{v} \sim \mathcal{CN}(0, \beta_r \bar{\mathbf{v}}^H\bar{\mathbf{R}}\bar{\mathbf{v}}+\beta_{d}),
\end{align}
where $\bar{\mathbf{v}}=(\mathbf{v}^T)^H$. The sum-rate scaling for this setting under a deterministic beamforming scheme is provided below.

\begin{theorem}\label{Lem:Cor_Rayleigh}
For correlated Rayleigh fading, the sum-rate in \eqref{AR2} under IRS-assisted OBF scales as \vspace{-.06in}
\begin{align}
\label{AR7}
&R^{(K)}=\log_2(1+ \frac{P}{\sigma^2}(\beta_r \alpha^2 \zeta +\beta_d)\log K),
\end{align}
as $K$ grows large, where $\zeta=\sum\limits_{j=1}^{N-1} \lambda_j |(e^{j \angle\mathbf{u}_{N}})^H \mathbf{u}_j|^2+\lambda_{N}|\sum_{i=1}^N |\mathbf{u}_{N}(i)||^2$, $\lambda_1< \dots < \lambda_N$ are the eigenvalues of  $\bar{\mathbf{R}}$ and $\mathbf{u}_j$, $j=1,\dots, N$ are the associated eigenvectors. This scaling is achieved  by a deterministic design for $\mathbf{v}$ satisfying $|v_n|=\alpha$ $\forall n$, given as \vspace{-.05in}
\begin{align}
\label{design}
&\bar{\mathbf{v}}=\alpha e^{j \angle\mathbf{u}_{N}}.
\end{align}
\end{theorem}
\begin{IEEEproof} 
See Appendix \ref{Proof_Cor_Rayleigh}.
\end{IEEEproof}
The result yields the following important corollaries.

\begin{corollary}\label{Corr:Rayleigh1}
Under independent Rayleigh fading, i.e. $\mathbf{R}=\mathbf{I}_N$, the sum-capacity for any $\mathbf{v}$ scales as \vspace{-.06in}
\begin{align}
&R^{(K)}=\log_2 (1+ \frac{P}{\sigma^2}(\beta_r \alpha^2 N   +\beta_d)\log K).
\end{align}
\end{corollary}
\begin{IEEEproof} 
The proof follows from \eqref{AR4} using $\bar{\mathbf{v}}^H \bar{\mathbf{v}}=\alpha^2N$. It also follows from \eqref{AR7} by noting that $\lambda_j=1$, $\forall j$ and using the standard basis vectors as the eigenvectors.
\end{IEEEproof}

Therefore IRS-assisted OBF yields $N\alpha^2 \frac{\beta_r}{\beta_d}+1$ gain in the SNR of strongest user as compared to the system without IRS. 

\begin{corollary}\label{Corr:Rayleigh2}
Under completely correlated Rayleigh fading, i.e. $\mathbf{R}=\mathbf{a}\mathbf{a}^H$ where $\mathbf{a}$ is the array response vector, the sum-rate in \eqref{AR7} under the design in \eqref{design} scales as \vspace{-.07in}
\begin{align}
&R^{(K)}=\log_2 (1+ \frac{P}{\sigma^2}(\beta_r \alpha^2 N^2   +\beta_d)\log K).
\end{align}
\end{corollary}
\begin{IEEEproof} 
The proof follows from writing $\bar{\mathbf{v}}^H \bar{\mathbf{R}}\bar{\mathbf{v}}=\bar{\mathbf{v}}^H (\mathbf{h}_1 \circ \mathbf{a}) (\mathbf{h}_1 \circ \mathbf{a})^H \bar{\mathbf{v}}$. The only non-zero eigenvalue $\lambda_N=N$ and the associated eigenvector is $\mathbf{u}_N=\frac{1}{\sqrt{N}}\mathbf{h}_1\circ \mathbf{a}$.
\end{IEEEproof}

The sum-rate in \eqref{AR7} therefore lies in the interval $\log_2 (1+ \frac{P}{\sigma^2}(\beta_r \alpha^2 N  +\beta_d)\log K) \leq R^{(K)}\leq \log_2 (1+ \frac{P}{\sigma^2}(\beta_r \alpha^2 N^2   +\beta_d)\log K)$. Interestingly, we see that by exploiting the eigenvalue decomposition of $\bar{\mathbf{R}}$ to design $\mathbf{v}$, IRS-assisted OBF performs better under correlated Rayleigh fading than it does under independent Rayleigh fading, with approximately a factor of $N$ gain in the SNR of the strongest user under completely correlated fading.   The deterministic design in \eqref{design} depends only on the IRS correlation matrix, which is well-known to vary very slowly as compared to the fast fading process and can be computed after several coherence intervals using only statistical information. Therefore the IRS does not require instantaneous CSI to achieve the sum-rate in \eqref{AR7}.

\begin{remark}
The results in this paper can be extended to the multiple-input single-output (MISO) BC under the random beamforming (RBF) scheme discussed in \cite{tareq, TDMA1}, where the BS sends multiple random orthonormal beams in each coherence interval and on each beam schedules the user with the highest signal-to-interference-plus-noise ratio (SINR), thereby exploiting the MU-diversity effect.  Under RBF at the BS, we can study the effect of introducing an IRS employing random phase rotations  on the MU-diversity gain and sum-rate scaling of the MISO BC. This paper provides important fundamental analysis to make this extension in future works.
\end{remark}

\vspace{-.16in}
\section{Simulations}\label{Sec:Simulations}
\vspace{-.03in}

Using $(x,y)$ coordinates (in meters), the BS and IRS are deployed at $(0,0)$ and $(0,50)$ respectively, and the users are uniformly distributed in the region $(x,y)\in[-30,30]\times[50,130]$. We set $P=1W$, $\sigma^2=-80$\rm{dBm}, $\alpha=1$ and assume $5$\rm{dBi} elements at the BS and IRS. The path loss in the IRS-assisted link $\beta_{r,k}$ is the product of the path loss in BS-IRS link and the path loss in IRS-user link with path loss exponents $2.2$ and $2.8$ respectively, while that for ${h}_{d,k}$ is $3.5$ \cite{LIS}. Penetration loss of $10$\rm{dB} is assumed for the direct link.

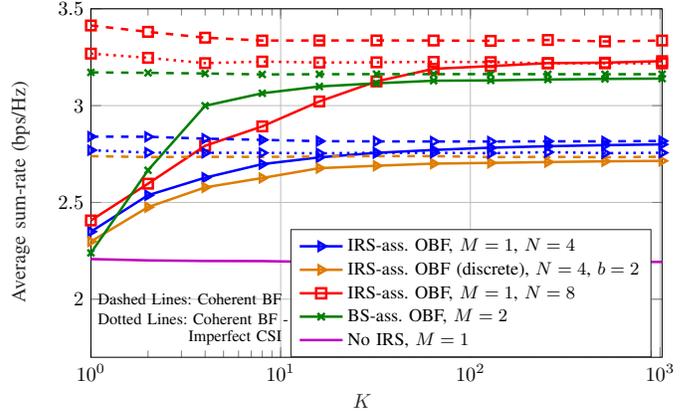
\begin{figure}[t]
\tikzset{every picture/.style={scale=.95}, every node/.style={scale=.8}}
%
%
\definecolor{mycolor1}{rgb}{0.00000,0.49804,0.00000}%
\definecolor{mycolor2}{rgb}{0.74902,0.00000,0.74902}%
\definecolor{mycolor3}{rgb}{0.87059,0.49020,0.00000}%
\definecolor{mycolor4}{rgb}{0.49412,0.18431,0.55686}%
\begin{tikzpicture}

\begin{axis}[%
width=.95\columnwidth,
height=.58\columnwidth,
scale only axis,
xmin=1,
xmax=1024,
xmode=log,
xlabel style={font=\color{white!15!black}},
xlabel={$K$},
ymin=1.7,
ymax=3.5,
ylabel style={font=\color{white!15!black}},
ylabel={Average sum-rate (bps/Hz)},
axis background/.style={fill=white},
xmajorgrids,
ymajorgrids,
legend style={at={(axis cs: 900,1.7)},anchor=south east,legend cell align=left,align=left,draw=white!15!black, /tikz/column 2/.style={
                column sep=5pt,
            }},
						]
\addplot [color=blue,  solid, line width=1.0pt,mark size=2.0pt,mark=triangle,mark options={solid, rotate=270, blue}]
  table[row sep=crcr]{%
1	2.34890920035452\\
2	2.53690548749093\\
4	2.62778660576967\\
8	2.69777373772299\\
16	2.73272619634136\\
32	2.75613301298736\\
64	2.77086615111116\\
128	2.78264578469676\\
256	2.79019676877592\\
512	2.79576769425277\\
1024	2.79973583110342\\
};
\addlegendentry{\small IRS-ass. OBF, $M=1$, $N=4$}

\addplot [color=mycolor3,solid,line width=1.0pt,mark size=2.0pt,mark=triangle,  mark options={solid, rotate=270, mycolor3}]
  table[row sep=crcr]{%
1	2.29804406671975\\
2	2.47543466591419\\
4	2.57761989752864\\
8	2.62614465917683\\
16	2.67692962737473\\
32	2.68933514954108\\
64	2.7003876262023\\
128	2.70417535190747\\
256	2.70839657916981\\
512	2.71233830900624\\
1024	2.71453569226109\\
};
\addlegendentry{\small IRS-ass. OBF (discrete), $N=4$, $b=2$}

\addplot [color=red,solid,line width=1.0pt,mark size=2.0pt,mark=square,  mark options={solid, rotate=270, red}]
  table[row sep=crcr]{%
1	2.40676648452379\\
2	2.59673731672421\\
4	2.7926110090152\\
8	2.89267821650745\\
16	3.021473038819\\
32	3.12492351616079\\
64	3.19019942171662\\
128	3.20414678125789\\
256	3.21952383134435\\
512	3.22182051571433\\
1024	3.22975483909817\\
};
\addlegendentry{\small IRS-ass. OBF, $M=1$, $N=8$}

\addplot [color=mycolor1,solid,line width=1.0pt,mark size=2.0pt,mark=x, mark options={solid, rotate=270, mycolor1}]
  table[row sep=crcr]{%
1	2.23906934167145\\
2	2.66600004168688\\
4	2.9991318114663\\
8	3.0638340799376\\
16	3.09886860390365\\
32	3.11525545300872\\
64	3.12859503967045\\
128	3.13030201190182\\
256	3.13485377066426\\
512	3.13795588998234\\
1024	3.13955122722443\\
};
\addlegendentry{\small BS-ass. OBF, $M=2$}

\addplot [color=mycolor2, solid,  line width=1.0pt]
  table[row sep=crcr]{%
1	2.20731522623469\\
2	2.20049195398683\\
4	2.19741678933825\\
8	2.19683980284145\\
16	2.19218244650222\\
32	2.19580009580535\\
64	2.19208315390694\\
128	2.19465820264938\\
256	2.19492637234315\\
512	2.19138181967157\\
1024	2.1926997453105\\
};
\addlegendentry{\small No IRS, $M=1$}

\addplot [color=mycolor1,  dashed,  line width=1.0pt,mark size=2.0pt,mark=x,mark options={solid, rotate=270, mycolor1}]
  table[row sep=crcr]{%
1	3.17169217659341\\
2	3.16923043194736\\
4	3.16597498982422\\
8	3.16113658541571\\
16	3.16219794022603\\
32	3.1625959775601\\
64	3.16255057364574\\
128	3.1627353780895\\
256	3.16229505035351\\
512	3.16215195870911\\
1024	3.16230783591055\\
};

\addplot [color=blue, dashed, line width=1.0pt,mark size=2.0pt,mark=triangle, mark options={solid, rotate=270, blue}]
  table[row sep=crcr]{%
1	2.83990623904553\\
2	2.83905739344412\\
4	2.82931195112149\\
8	2.82311628272672\\
16	2.81627396785769\\
32	2.81525404753187\\
64	2.81472465341783\\
128	2.81454121387336\\
256	2.815620032802\\
512	2.81535248604876\\
1024	2.81733360407016\\
};

\addplot [color=red, dashed, line width=1.0pt,mark size=2.0pt,mark=square,  mark options={solid, rotate=270, red}]
  table[row sep=crcr]{%
1	3.41388210146256\\
2	3.38103676747899\\
4	3.35096722328431\\
8	3.33605736685202\\
16	3.33630343512987\\
32	3.3365944529118\\
64	3.33621025295455\\
128	3.33464595540405\\
256	3.33887203295269\\
512	3.33167959073529\\
1024	3.33618563912197\\
};

\addplot [color=mycolor3, dashed, line width=1.0pt,mark size=2.0pt, mark=traingle, mark options={solid, mycolor3}]
  table[row sep=crcr]{%
1	2.73903610108788\\
2	2.73425188066326\\
4	2.73540846733168\\
8	2.73520721438951\\
16	2.73666913706946\\
32	2.73934777519984\\
64	2.73912450607615\\
128	2.73416162193935\\
256	2.73527839146031\\
512	2.73357936797953\\
1024	2.73581278131009\\
};

\addplot [color=blue, dotted, line width=1.0pt,mark size=2.0pt, mark=triangle, mark options={solid, rotate=270, blue}]
  table[row sep=crcr]{%
1	2.76966941964123\\
2	2.75718458540107\\
4	2.75622172685429\\
8	2.75600456498578\\
16	2.7534956638577\\
32	2.75626379016492\\
64	2.75614986387739\\
128	2.75425792185547\\
256	2.75958734097694\\
512	2.75437737254101\\
1024	2.75643578350097\\
};

\addplot [color=red, dotted, line width=1.0pt,mark size=2.0pt, mark=square, mark options={solid, red}]
  table[row sep=crcr]{%
1	3.26867377704746\\
2	3.24700499113695\\
4	3.21933459091206\\
8	3.22755684293018\\
16	3.22257888374179\\
32	3.22370206502822\\
64	3.22618093029475\\
128	3.22474629833833\\
256	3.21879827526651\\
512	3.21857439313948\\
1024	3.21832965294291\\
};

\node at (axis cs: 1,2.0) [anchor=west] {\footnotesize Dashed Lines: Coherent BF};
\node at (axis cs: 1,1.9) [anchor=west] {\footnotesize Dotted Lines: Coherent BF -};
\node at (axis cs: 3,1.8) [anchor=west] {\footnotesize Imperfect CSI};


\end{axis}
\end{tikzpicture}%
\caption{Sum-rate performance under slow-fading.}
\label{Fig1}
\end{figure}

In the first result, the slow fading realizations of $\mathbf{h}_{2,k}$ and ${h}_{d,k}$ are generated as i.i.d. Rayleigh distributed and stay constant over the latency time-scale. We verify Theorem 2  in Fig. \ref{Fig1} by plotting the average sum-rate  in \eqref{AR}  against $K$ for $N=4$ and $N=8$, with the IRS phases drawn randomly from the distribution specified by \eqref{eq_imp} for IRS-assisted OBF. The performance under coherent beamforming (the eventual limit in \eqref{Thm1_1}) is also plotted. We see the average sum-rate increase with $K$ under IRS-assisted OBF while it stays constant when there is no IRS. For $N=4$, the sum-rate of the proposed IRS-assisted OBF scheme almost approaches the coherent beamforming performance (that requires full CSI) for $K=128$ users. The convergence slows down for large $N$ because the probability of mismatch between the random phase-shifts and beamforming configuration of the scheduled user in \eqref{eq_imp} increases \cite{slowfad}. However, the sum-rate gap reduces as $K$ grows large and the sum-rate under IRS-assisted OBF will eventually approach the limit as stated in Theorem 2.  

The number of  passive IRS elements required to outperform BS-assisted OBF scheme from \cite{dumb}, that uses an $M$-antenna BS to yield MU-diversity gains, is higher to overcome the double path loss.  However, IRS-assisted OBF relies on a single-antenna BS, making it an energy-efficient alternative. 

In addition to coherent beamforming under perfect CSI, we also consider coherent beamforming under the more practical imperfect CSI scenario as the second benchmark. We use the imperfect CSI model $\hat{\mathbf{h}}_{2,k}=\sqrt{1-\epsilon^2}\mathbf{h}_{2,k}+\epsilon_k \Delta_k$, where $\hat{\mathbf{h}}_{2,k}$ is the estimated channel (also constant over latency time-scale) and $\Delta_k$ is the channel error vector. The maximum achievable rate of each user is determined by solving $\underset{|v_n|=1}{\text{max}}  |\hat{h}_k|^2$ to find $\mathbf{v}$ similar to \eqref{RBF}. The resulting performance is plotted in Fig. \ref{Fig1} for  $\epsilon_k=0.2$. We see that for $N=4$ and $N=8$, OBF outperforms coherent beamforming under imperfect CSI for $K>32$ and $K>256$ users respectively and eventually achieves the coherent beamforming under perfect CSI upper bound. Moreover, it is important to remark that in contrast to OBF, coherent and robust beamforming \cite{robust} schemes require the IRS phase shifts to be computed based on the knowledge of $\hat{\mathbf{h}}_{2,k}$ \cite{robust}.

In practice the IRS elements can implement a finite number of phase-shifts depending on the the number of bits $b$ representing their resolution. We show the performance loss caused by quantizing the continuous phase shifts in \eqref{conf} and \eqref{eq_imp} to the nearest discrete values for $N=4$ in Fig. \ref{Fig1}. While there is a performance loss, the convergence in Theorem 2 still holds.

\begin{figure}[t]
\tikzset{every picture/.style={scale=.95}, every node/.style={scale=.8}}
%
%
\definecolor{mycolor1}{rgb}{0.00000,0.49804,0.00000}%
\definecolor{mycolor2}{rgb}{0.74902,0.00000,0.74902}%
\begin{tikzpicture}

\begin{axis}[%
width=.95\columnwidth,
height=.55\columnwidth,
scale only axis,
xmin=0,
xmax=1,
xlabel style={font=\color{white!15!black}},
xlabel={$\eta$},
ymin=4.2,
ymax=11,
ylabel style={font=\color{white!15!black}},
ylabel={Average sum-rate (bps/Hz)},
axis background/.style={fill=white},
xmajorgrids,
ymajorgrids,
legend style={at={(axis cs: 0,11)},anchor=north west,legend cell align=left,align=left,draw=white!15!black}
]
\addplot [color=red, mark=x,line width=1.0pt,mark size=2pt, mark options={solid, red}]
  table[row sep=crcr]{%
0	5.76501358518347\\
0.2	5.78119153004759\\
0.4	5.75003129935904\\
0.6	5.76017396976201\\
0.8	5.77442912304879\\
1	5.76392233943767\\
};
\addlegendentry{\small Ray., IRS-ass. OBF}

\addplot [color=red, dashed, mark=x, line width=1.0pt,mark size=2pt,mark options={solid, red}]
  table[row sep=crcr]{%
0	5.64789580678189\\
0.2	5.64789580678189\\
0.4	5.64789580678189\\
0.6	5.64789580678189\\
0.8	5.64789580678189\\
1	5.64789580678189\\
};
\addlegendentry{\small Corollary \ref{Corr:Rayleigh1}}

\addplot [color=red, mark=triangle, line width=1.0pt,mark size=2pt, mark options={solid, rotate=270, red}]
  table[row sep=crcr]{%
0	5.76501358518347\\
0.2	5.90409793784085\\
0.4	6.03944656305224\\
0.6	6.26830875554038\\
0.8	6.61607258784009\\
1	7.06206322692696\\
};
\addlegendentry{\small Corr. Ray., IRS-ass. OBF (Det. $\mathbf{v}$)}

\addplot [color=red, dashed, mark=triangle, line width=1.0pt,mark size=2pt,mark options={solid, rotate=270, red}]
  table[row sep=crcr]{%
0	5.64789580678189\\
0.2	5.77677264699722\\
0.4	5.94783730288633\\
0.6	6.18001811336699\\
0.8	6.50018924212248\\
1	6.94143219554422\\
};
\addlegendentry{\small Theorem \ref{Lem:Cor_Rayleigh}}

\addplot [color=black, mark=diamond, line width=1.0pt,mark size=2.4pt, mark options={solid, black}]
  table[row sep=crcr]{%
0	5.42565739286403\\
0.2	5.4225731258724\\
0.4	5.42967092021538\\
0.6	5.41773280575127\\
0.8	5.41739169698651\\
1	5.41663148962325\\
};
\addlegendentry{\small No IRS, $M=1$}

\addplot [color=mycolor2, dotted, mark=o,line width=1.0pt,mark size=1.5pt, mark options={solid, mycolor2}]
  table[row sep=crcr]{%
0	5.42154721620927\\
0.2	5.40991476646473\\
0.4	5.43388325073227\\
0.6	5.40820252922989\\
0.8	5.43037472155751\\
1	5.42702049620466\\
};
\addlegendentry{\small Ray., BS-ass. OBF, $M=2$}

\addplot [color=mycolor1, mark=x,  line width=1.0pt,mark size=2pt,mark options={solid, mycolor1}]
  table[row sep=crcr]{%
0	6.64350324046651\\
0.2	6.66367639375773\\
0.4	6.65771736885294\\
0.6	6.65182743662865\\
0.8	6.64709785516175\\
1	6.64519883847132\\
};

\addplot [color=mycolor1, dashed, mark=x,  line width=1.0pt,mark size=2pt,mark options={solid, mycolor1}]
  table[row sep=crcr]{%
0	6.54565000944445\\
0.2	6.54565000944445\\
0.4	6.54565000944445\\
0.6	6.54565000944445\\
0.8	6.54565000944445\\
1	6.54565000944445\\
};

\addplot [color=mycolor1, mark=triangle,  line width=1.0pt,mark size=2pt,mark options={solid, rotate=270, mycolor1}]
  table[row sep=crcr]{%
0	6.64350324046651\\
0.2	6.98049472266838\\
0.4	7.38260935942602\\
0.6	7.92279620330693\\
0.8	8.81081108392966\\
1	10.7405153734224\\
};

\addplot [color=mycolor1, dashed, mark=triangle,  line width=1.0pt,mark size=2pt,mark options={solid, rotate=270, mycolor1}]
  table[row sep=crcr]{%
0	6.54565000944445\\
0.2	6.87166409163869\\
0.4	7.28612774178504\\
0.6	7.85425039731775\\
0.8	8.74723805048736\\
1	10.6203304596254\\
};
\node at (axis cs: .05,4.4) [anchor=west] {\footnotesize Green: $M=1$, $N=32$};
\node at (axis cs: .05,4.7) [anchor=west] {\footnotesize Red: $M=1$, $N=8$};

\end{axis}
\end{tikzpicture}%
\caption{Sum-rate under independent Rayleigh (Ray.) and correlated (Corr.) Rayleigh channels for $K=256$.}
\label{Fig2}
\end{figure}
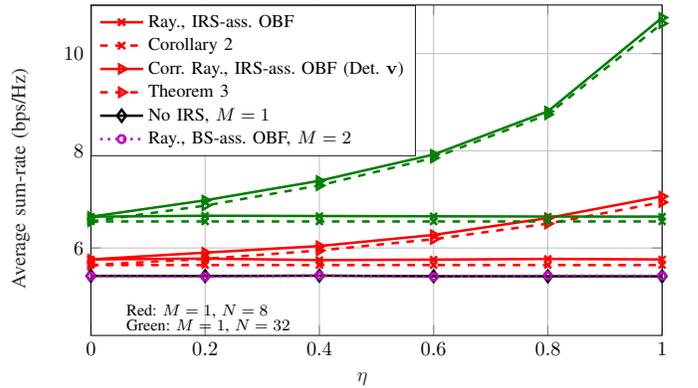

In Fig. \ref{Fig2}, we study the average sum-rate performance in \eqref{AR}  under independent and correlated Rayleigh fading channels.  The covariance matrix is set as $[\mathbf{R}]_{i,j}=\eta^{|i-j|}$, $i,j=1,\dots, N$. Under independent Rayleigh fading with uniformly distributed IRS phases, we see a sum-rate gain of approximately $\log_2 (N \alpha^2 \frac{\beta_{r}}{\beta_{d}} +1)$ as compared to the system without IRS, in line with Corollary \ref{Corr:Rayleigh1}, which is also plotted.

 Under correlated Rayleigh fading and deterministic (Det.) design for $\mathbf{v}$ in \eqref{design}, we plot both \eqref{AR} as well as the theoretical scaling law in Theorem \ref{Lem:Cor_Rayleigh} and show a good match. Note that the match will become more accurate for larger $K$ as promised by Theorem \ref{Lem:Cor_Rayleigh}. The performance improves as $\eta$ increases from $0$ to $1$, confirming that with the design in \eqref{design} that only depends on the channel covariance matrix, IRS-assisted OBF performs better under correlated Rayleigh fading than under independent Rayleigh fading with the sum-rate gain becoming $\log_2 (N^2 \alpha^2 \frac{\beta_{r}}{\beta_{d}}+1)$  over a system without the IRS as $\eta$  approaches $1$.  Finally, we observe that having multiple antennas at the BS under BS-assisted OBF \cite{dumb} yields no gain under independent Rayleigh fading. 

\vspace{-.1in}
\section{Conclusion}
\vspace{-.02in}
We have proposed an IRS-assisted OBF scheme for a SISO BC, in which the IRS elements employ either random or deterministic phase rotations to increase the MU-diversity gains while the BS employs PF scheduling. Without requiring instantaneous CSI to design the IRS phases, we show the average sum-rate under the proposed scheme  in the slow-fading environment to approach coherent beamforming performance as the number of users increases. Further,  we show this technique to improve the sum-rate gain in independent and correlated Rayleigh fast-fading  environments as well.  

\vspace{-.1in}
\appendices

\section{Proof of Theorem \ref{Thm:slow}}
\label{Proof_Thm_slow}
\vspace{-.02in}
Denote the discrete slow fading states of $\mathbf{h}_{2,k}$ and $h_{d,k}$ as $\mathbf{h}_{2,p}$, $p\in \mathcal{P}=\{1,\dots, P\}$ and $h_{d,q}$, $q\in \mathcal{Q}=\{1,\dots, Q\}$ respectively and the discrete state pairs as $(\mathbf{h}_{2,p_{p\in \mathcal{P}}},h_{d,q_{q\in \mathcal{Q}}})_{j}$, $j=1,\dots, M$ where $M=PQ$. Also denote  the probability of user being in state $j$ as $p_j$.  A discrete number of fading states is assumed to minimize the technicality of the proof \cite{dumb}. The maximum achievable rate in state $j$ is $R_j^{BF}$ given in \eqref{RBF}. The theorem implies that the joint stationary distribution of IRS phase variation process also has probability $p_j$ on state $(\theta^{BF}_{1,j}, \dots, \theta^{BF}_{N,j})$, and by ergodicity, this is the long-term fraction of time the process spends in this state. Denote the fraction of users in class $j$ as $c_j^{(K)}$. We will have $\sum_{j=1}^M c_j^{(K)}=1$ and \vspace{-.08in}
\begin{align}
\label{333}
&\underset{K\rightarrow \infty}{\text{lim}}c_j^{(K)}=p_j, j=1,\dots, M.
\end{align}

The proof follows by developing lower and upper bound on the average throughput of the users. The lower bound is obtained using a simple scheduling algorithm that schedules a user in class $j$  only when the IRS phases are in the beamforming configuration $(\theta^{BF}_{1,j}, \dots, \theta^{BF}_{N,j})$. Such a user exists almost surely when $K$ is large. This way, the long-term average rate of a user in class $j$ is $\frac{p_j R_j^{BF}}{c_j^{(K)} K}$. Using \cite[Lemma 4]{dumb} that PF algorithm maximizes $\sum_{k=1}^K \log T_k^{(K)}$ almost surely among the class of all schedulers, we obtain \vspace{-.08in}
\begin{align}
\label{222}
\sum_{k=1}^K \log T_k^{(K)} \geq \sum_{j=1}^M c_j^{(K)} K \log\left(\frac{p_j R_j^{BF} }{c_j^{(K)} K}  \right).
\end{align}

Now consider the PF algorithm and denote by $d_j^{(K)}$ the fraction of time it schedules users in state $j$. We have the following upper bound: \vspace{-.06in}
\begin{align}
\label{111}
\sum_{k=1}^K \log T_k^{(K)} \leq \sum_{j=1}^M c_j^{(K)} K \log\left(\frac{d_j^{(K)} R_j^{BF} }{c_j^{(K)} K}  \right)
\end{align}

The proof then follows from combining \eqref{222} and \eqref{111} and using \eqref{333} to prove that $\underset{K\rightarrow \infty}{\text{lim}} d_j^{(K)} =p_j$. Using this result in \eqref{111}, the average throughput of any user $k$ in class $j$ under PF scheduling will satisfy  $\underset{K\rightarrow \infty}{\text{lim inf}} K  T_k^{(K)} \leq R_j^{BF}$. Combining this with the lower bound in \eqref{222} and using \eqref{333}, we can complete the proof (more details in \cite{dumb} that studies BS-assisted OBF).

\vspace{-.1in}
\section{Proof of Theorem \ref{Lem:Cor_Rayleigh}}
\label{Proof_Cor_Rayleigh}
\vspace{-.05in}
The derivation will utilize \cite[Lemma 2]{dumb}, which states that the maximum of $K$ i.i.d. RVs with pdf $f_{X}(x)$ and cdf $F_{X}(x)$  grows like $l_K$ as $K\to\infty$ if $\underset{x\rightarrow \infty}{\lim} g(x)=\frac{1-F_{X}(x)}{f_{X}(x)}=c>0$, where $l_K$ is obtained as the solution of $F_X(l_{K})=1-\frac{1}{K}$.

Given $\mathbf{v}$ and using \eqref{hh}, $\gamma_k$ will follow exponential distribution with parameter $\frac{\sigma^2}{P(\beta_r \bar{\mathbf{v}}^H\bar{\mathbf{R}}\bar{\mathbf{v}}+\beta_d)}$, resulting in $F_{\gamma}(x)= 1-e^{-\frac{\sigma^2 x}{P(\beta_r \bar{\mathbf{v}}^H\bar{\mathbf{R}}\bar{\mathbf{v}}+\beta_d)}}$ and $f_{\gamma}(x)=\frac{\sigma^2}{P(\bar{\mathbf{v}}^H\bar{\mathbf{R}}\bar{\mathbf{v}} \beta_r+\beta_d)} e^{-\frac{\sigma^2 x}{P(\bar{\mathbf{v}}^H\bar{\mathbf{R}}\bar{\mathbf{v}} \beta_r+\beta_d)}}$.  We now obtain $g(x)=\frac{1-F_{\gamma}(x)}{f_{\gamma}(x)} =\frac{P}{\sigma^2}(\beta_r \bar{\mathbf{v}}^H\bar{\mathbf{R}}\bar{\mathbf{v}} +\beta_d)>0$,  satisfying the condition.  Solving for $l_K$, we have $e^{-\frac{\sigma^2 l_K}{P(\beta_r \bar{\mathbf{v}}^H\bar{\mathbf{R}}\bar{\mathbf{v}}+\beta_d)}}=\frac{1}{K}$ and therefore $l_{K}=\frac{P}{\sigma^2}(\beta_r \bar{\mathbf{v}}^H\bar{\mathbf{R}}\bar{\mathbf{v}}+\beta_d) \log K$. Substituting $l_K$ for $\underset{k}{\max} \gamma_{k}$ in \eqref{AR2}, we obtain \vspace{-.1in}
\small 
\begin{align}
\label{AR4}
&R^{(K)}=\underset{f(\bar{\mathbf{v}})}{\text{max } }\mathbb{E}_{\bar{\mathbf{v}}}[\log_2(1+ \frac{P}{\sigma^2}(\beta_r \bar{\mathbf{v}}^H \bar{\mathbf{R}} \bar{\mathbf{v}} +\beta_d)\log K)].
\end{align}  \normalsize
Consider a deterministic OBF scheme, where $\bar{\mathbf{v}}$ is fixed over all $t$. Then \eqref{AR4} can be written as $R^{(K)}=\log_2(1+ \frac{P}{\sigma^2}(\underset{\bar{\mathbf{v}}}{\text{max }} \bar{\mathbf{v}}^H \mathbf{U} \boldsymbol{\Lambda} \mathbf{U}^H \bar{\mathbf{v}} \beta_R  +\beta_d)\log K)$, where $\mathbf{U} \boldsymbol{\Lambda} \mathbf{U}^H $ is the eigenvalue decomposition of $\bar{\mathbf{R}}$. To solve this maximization, we first relax the constraint $|\bar{v}_n|=\alpha$ as $\|\tilde{\mathbf{v}} \|^2=\alpha^2 N$. The maximum value of $\tilde{\mathbf{v}}^H \mathbf{U} \boldsymbol{\Lambda} \mathbf{U}^H \tilde{\mathbf{v}}$  is achieved when $\tilde{\mathbf{v}}=\sqrt{N}\alpha \mathbf{u}_{N}$, where $\mathbf{u}_{N}$ is the eigen-vector corresponding to the maximum eigenvalue $\lambda_{N}$ of $\bar{\mathbf{R}}$. The corresponding $\bar{\mathbf{v}}$ satisfying $|\bar{v}_n|=\alpha$, $\forall n$ is obtained as the solution of $\underset{\bar{\mathbf{v}}}{\text{min}}|\bar{\mathbf{v}}-\tilde{\mathbf{v}}|^2$ and is given by \eqref{design} \cite{LIS}. Plugging \eqref{design} in \eqref{AR4} yields \eqref{AR7}.

\vspace{-.07in}

\bibliographystyle{IEEEtran}
\bibliography{bib}

\begin{thebibliography}{10}
\providecommand{\url}[1]{#1}
\csname url@samestyle\endcsname
\providecommand{\newblock}{\relax}
\providecommand{\bibinfo}[2]{#2}
\providecommand{\BIBentrySTDinterwordspacing}{\spaceskip=0pt\relax}
\providecommand{\BIBentryALTinterwordstretchfactor}{4}
\providecommand{\BIBentryALTinterwordspacing}{\spaceskip=\fontdimen2\font plus
\BIBentryALTinterwordstretchfactor\fontdimen3\font minus
  \fontdimen4\font\relax}
\providecommand{\BIBforeignlanguage}[2]{{%
\expandafter\ifx\csname l@#1\endcsname\relax
\typeout{** WARNING: IEEEtran.bst: No hyphenation pattern has been}%
\typeout{** loaded for the language `#1'. Using the pattern for}%
\typeout{** the default language instead.}%
\else
\language=\csname l@#1\endcsname
\fi
#2}}
\providecommand{\BIBdecl}{\relax}
\BIBdecl

\bibitem{huang}
C.~{Huang} \emph{et~al.}, ``Reconfigurable intelligent surfaces for energy
  efficiency in wireless communication,'' \emph{IEEE Trans. Wireless Commun.},
  vol.~18, no.~8, pp. 4157--4170, Aug 2019.

\bibitem{LIS}
Q.~{Wu} and R.~{Zhang}, ``Intelligent reflecting surface enhanced wireless
  network via joint active and passive beamforming,'' \emph{IEEE Trans.
  Wireless Commun}, vol.~18, no.~11, pp. 5394--5409, Nov 2019.

\bibitem{robust}
G.~{Zhou}, C.~{Pan}, H.~{Ren}, K.~{Wang}, M.~{Di Renzo}, and A.~{Nallanathan},
  ``Robust beamforming design for intelligent reflecting surface aided miso
  communication systems,'' \emph{IEEE Wireless Commun. Lett.}, pp. 1--1, 2020.

\bibitem{annie}
Q.~{Nadeem} \emph{et~al.}, ``Asymptotic max-min {SINR} analysis of
  reconfigurable intelligent surface assisted {MISO} systems,'' \emph{IEEE
  Trans. Wireless Commun.}, pp. 1--1, 2020.

\bibitem{annie_OJ}
------, ``Intelligent reflecting surface-assisted multi-user {MISO}
  communication: Channel estimation and beamforming design,'' \emph{IEEE Open
  Journal of the Communications Society}, vol.~1, pp. 661--680, 2020.

\bibitem{broadband}
Z.~{Wan}, Z.~{Gao}, and M.-S. {Alouini}, ``{Broadband Channel Estimation for
  Intelligent Reflecting Surface Aided mmWave Massive {MIMO} Systems},'' in
  \emph{IEEE International Conference on Communications (ICC)}, 2020.

\bibitem{TDMA1}
M.~{Sharif} and B.~{Hassibi}, ``A comparison of time-sharing, {DPC}, and
  beamforming for {MIMO} broadcast channels with many users,'' \emph{IEEE
  Trans. Commun.}, vol.~55, no.~1, pp. 11--15, Jan 2007.

\bibitem{dumb}
P.~{Viswanath}, D.~N.~C. {Tse}, and R.~{Laroia}, ``Opportunistic beamforming
  using dumb antennas,'' \emph{IEEE Trans. Information Theory}, vol.~48, no.~6,
  pp. 1277--1294, June 2002.

\bibitem{slowfad}
{Jaehak Chung} \emph{et~al.}, ``A random beamforming technique in {MIMO}
  systems exploiting multiuser diversity,'' \emph{IEEE Journal on Selected
  Areas in Communications}, vol.~21, no.~5, pp. 848--855, 2003.

\bibitem{tareq}
T.~{Al-Naffouri}, M.~{Sharif}, and B.~{Hassibi}, ``How much does transmit
  correlation affect the sum-rate scaling of {MIMO} gaussian broadcast
  channels?'' \emph{IEEE Trans. Commun.}, vol.~57, pp. 562--572, Feb. 2009.

\bibitem{rotate1}
C.~{Psomas} and I.~{Krikidis}, ``{Low-Complexity Random Rotation-based Schemes
  for Intelligent Reflecting Surfaces},'' \emph{arXiv e-prints}, p.
  arXiv:1912.10347, Dec. 2019.

\bibitem{dumb_me}
Q.-U.-A. {Nadeem}, A.~{Chaaban}, and M.~{Debbah}, ``Reconfigurable surface
  assisted multi-user opportunistic beamforming,'' in \emph{IEEE International
  Symposium on Information Theory (ISIT)}, 2020.

\bibitem{rotate}
R.~{Pedarsani}, O.~{Lévêque}, and S.~{Yang}, ``On the {DMT} optimality of
  time-varying distributed rotation over slow fading relay channels,''
  \emph{IEEE Trans. Wireless Commun.}, vol.~14, no.~1, pp. 421--434, 2015.

\end{thebibliography}
\end{document}